\documentclass[conference]{IEEEtran}
\IEEEoverridecommandlockouts
\usepackage[utf8]{inputenc}
\usepackage{cite}
\usepackage{overpic}
\usepackage{amsmath,amssymb,amsfonts}
\usepackage{bm}
\usepackage{graphicx}
\usepackage{textcomp}
\usepackage{xcolor}
\usepackage{float}
\usepackage{amsthm}
\usepackage{graphicx}
\usepackage{epstopdf}
\usepackage{amsmath,bm,bbm}
\usepackage{amsfonts}
\usepackage{amssymb}
\usepackage{color}
\usepackage{multirow}
\usepackage{multicol}
\usepackage{soul,xcolor}
\usepackage{setspace}

\usepackage{longtable}
\usepackage{mathtools}
\usepackage{subfig}
\usepackage{tabulary}
\usepackage{epsfig}
\usepackage{caption}
\usepackage{booktabs}
\usepackage{blindtext}
\usepackage{adjustbox}
\usepackage{hyperref}
\usepackage{dirtytalk}
\usepackage{comment} 
\usepackage{tikz}
\usetikzlibrary{positioning} 
\usepackage{algorithm}
\usepackage{algpseudocode}

\definecolor{mypurple}{HTML}{9933FF}
\definecolor{mygreen}{HTML}{009900}

\theoremstyle{plain}

\newcommand{\vect}[1]{\mathbf{#1}}

\newcommand{\minimize}[1]{{\underset{{#1}}{\mathrm{minimize}}}}

\def\diag{\mathrm{diag}}

\def\kron{\otimes}

\def\Htran{\mbox{\tiny $\mathrm{H}$}}
\def\Ttran{\mbox{\tiny $\mathrm{T}$}}
\def\CN{\mathcal{N}_{\mathbb{C}}} 
\def\imagunit{\mathsf{j}} 
\begin{document}
\bstctlcite{IEEEexample:BSTcontrol}
\makeatletter
\newcommand*{\rom}[1]{\expandafter\@slowromancap\romannumeral #1@}
\makeatother

\title{Constant-Envelope Quantized Precoding with Power Control for Cell-Free Massive MIMO-OFDM \vspace{-0.4cm}  }

\author{\IEEEauthorblockN{\"Ozlem Tu\u{g}fe Demir}
\IEEEauthorblockA{\textit{Department of Electrical and Electronics Engineering} \\
\textit{Bilkent University}\\
Ankara, Turkiye \\
E-mail: ozlemtugfedemir@bilkent.edu.tr} 
\and
\IEEEauthorblockN{Salih Gümüsbuğa}
\IEEEauthorblockA{\textit{Department of Electrical and Electronics Engineering} \\
\textit{TOBB University of Economics and Technology}\\
Ankara, Turkiye \\
E-mail: s.gumusbuga@etu.edu.tr} 
\thanks{This work was carried out within the scope of the project 122C149 – Intelligent End-to-End Design of Energy-Efficient and Hardware Impairments-Aware Cell-Free Massive MIMO for Beyond 5G. \"O. T. Demir was supported by the 2232-B International Fellowship for Early Stage Researchers Programme funded by the Scientific and Technological Research Council of Türkiye (TÜBİTAK).
}}
\maketitle
\begin{abstract}
  Cell-free massive MIMO has matured into a key candidate technology for 6G and beyond, owing to its ability to provide nearly uniform service quality to many user equipments (UEs) over the same time–frequency resources. Unlike conventional cellular massive MIMO, the core idea is to distribute a large number of low-cost access points (APs) across the network and enable joint coherent transmission and reception. While early works largely assumed ideal hardware, hardware impairments become inevitable when APs are implemented with low-cost components. In this context, this paper investigates the adverse impact of low-resolution digital-to-analog converters (DACs) on the downlink performance of cell-free massive MIMO-OFDM systems. In contrast to prior studies that mainly quantify spectral-efficiency degradation under low-resolution DACs, we consider the design of quantized constant-envelope (CE) precoding, which additionally enables the use of highly power-efficient amplifiers. To the best of our knowledge, this is the first work on quantized CE precoding for cell-free massive MIMO-OFDM. Beyond adapting the classical maximum-antenna-power method, we propose a novel power-control strategy across APs that mitigates the detrimental effects of severely quantized transmitters by reducing the contribution of harmful APs. Simulation results demonstrate that the proposed power-control mechanism significantly improves the uncoded bit error rate performance. 
\end{abstract}

\section{Introduction}

The evolution toward large-scale antenna arrays has established massive MIMO (multiple-input multiple-output) as a fundamental enabler of modern cellular networks \cite{bjornson2019massive}. When the number of base station antennas greatly exceeds the number of user equipments (UEs), high spectral efficiency can be achieved through computationally efficient linear precoding strategies such as zero forcing (ZF), minimum mean squared error (MMSE), and regularized zero forcing (RZF) \cite{Larsson2014-ct}, \cite{Lu2014-ar}. Most of these designs, however, rely on the availability of high-resolution digital-to-analog converters (DACs). As antenna arrays continue to scale, the resulting hardware complexity, power consumption, and control overhead pose significant implementation challenges, which has stimulated interest in low-resolution and constant-envelope (CE) transmission architectures as more energy-efficient alternatives.

To overcome these limitations, quantized precoding has emerged as a viable framework for enabling hardware-efficient massive MIMO transmissions with coarse-resolution DACs. The squared-infinity norm Douglas–Rachford splitting (SQUID) algorithm introduced in \cite{Jacobsson2017a} demonstrated the effectiveness of nonlinear quantized precoding in multi-user scenarios and was compared against several benchmark schemes, including ZF, semidefinite relaxation (SDR), Wiener filtering (WF), and maximum ratio transmission (MRT). The impact of coarse quantization on WF was further investigated in \cite{Mezghani2009-fy} through uncoded bit error rate (BER) analysis, while power-constrained precoding formulations were studied in \cite{Yuan2020-hz}. Other works examined the performance of one-bit ZF relative to maximum-likelihood precoding \cite{Saxena2016-gs} and proposed low-complexity nonlinear precoders with competitive performance \cite{Park2019-sz}. 

CE precoding has also gained increasing attention due to its capability to generate zero peak-to-average power ratio (PAPR) discrete time-domain signals, which enables the use of highly efficient nonlinear power amplifiers in OFDM systems \cite{jacobsson2018nonlinear}. In particular, nonlinear phase-quantized CE precoding methods developed for massive multi-user MIMO-OFDM systems have shown that enforcing a CE constraint directly on the time-domain waveform can maintain reliable performance while improving overall hardware efficiency \cite{jacobsson2018nonlinear}.

Moving beyond co-located antenna arrays, cell-free massive MIMO distributes a large number of access points (APs), each equipped with a small number of antennas, across the service area to provide joint transmission to UEs. Owing to the ultra-dense deployment of APs, scalable and energy-efficient network operation requires low-cost hardware solutions. Although CE quantized transmission has been extensively investigated for conventional massive MIMO, its application to downlink cell-free architectures remains relatively unexplored. In our recent work on energy-efficient quantized precoding for cell-free systems \cite{gumusbuuga2025green}, we incorporated antenna deactivation through group-sparsity-aware design, showing that adaptive antenna usage can substantially reduce RF-chain activity while maintaining reliable performance.

In this paper, we go beyond antenna activation and instead investigate power control within CE quantized precoding for the downlink of cell-free massive MIMO, while extending the framework to OFDM-based transmission. The proposed method operates on time-domain CE waveforms and introduces AP-level power-control coefficients that adapt the transmit contribution of each distributed node. In contrast to conventional approaches that assume fixed transmit power, our design enables adaptive power scaling across APs while preserving the CE structure, leading to improved BER performance in cell-free massive MIMO systems.
\section{System Model}

We consider the downlink of a cell-free massive MIMO system employing OFDM transmission.
A set of $L$ distributed APs jointly serve $K$ single-antenna UEs that are randomly distributed in the coverage area.
Each AP is equipped with $N$ antennas, resulting in a total of $M = LN$ transmit antennas.
All APs are connected to a central processing unit (CPU) via ideal fronthaul links and operate in a time- and phase-synchronized manner.

We let $\vect{h}_{kl}[t]\in \mathbb{C}^{N}$ denote the $t$th tap of the frequency selective channel between UE $k$ and AP $l$ for $t=0,\ldots,T$. We assume perfect channel state information (CSI) at the APs and CPU to focus solely on the adverse impact of the low-resolution DACs. Concatenating all the channel vectors into a matrix, we obtain 
\begin{align}
    \vect{H}[t] = \begin{bmatrix} \vect{h}_{11}^{\Ttran}[t] & \cdots & \vect{h}_{1L}^{\Ttran}[t] \\ \vdots & \ddots & \vdots \\ \vect{h}_{K1}^{\Ttran}[t] & \cdots & \vect{h}_{KL}^{\Ttran}[t] \end{bmatrix} \in \mathbb{C}^{K\times M}.
\end{align}
 We let $\vect{x}[n]$ denote the quantized precoding signal after passing through DACs at discrete time instant $n=0,\ldots,S-1$ of an OFDM symbol, where $S>T$ is the number of samples per OFDM symbol, which is also equivalent to the number of subcarriers and discrete Fourier transform (DFT) size. The received signal at the UEs is given as
\begin{align}
\vect{y}[n] = \sum_{t=0}^T\vect{H}[t]\vect{x}[n-t]+\vect{w}[n]
\end{align}
where $\vect{w}[n]\sim \CN(\vect{0},\sigma^2\vect{I}_K)$ is the additive white Gaussian noise. We define
\(\vect{X} = [\vect{x}[0], \ldots, \vect{x}[S-1]]\), 
\(\vect{Y} = [\vect{y}[0], \ldots, \vect{y}[S-1]]\), and 
\(\vect{W} = [\vect{w}[0], \ldots, \vect{w}[S-1]]\). Denoting the scaled $S\times S$ DFT matrix (so that it is unitary) by $\vect{F}_S$, we obtain the frequency-domain signals in matrix form as
\begin{align}
\overline{\vect{X}} = \vect{X} \vect{F}_{S}, \quad 
\overline{\vect{Y}} = \vect{Y} \vect{F}_{S}, \quad 
\overline{\vect{W}} = \vect{W} \vect{F}_{S}. \quad 
\end{align}
Moreover, we define the frequency-domain channel at subcarrier $\nu$ as
\begin{align}
\overline{\vect{H}}[\nu] = \sum_{t=0}^{T} \vect{H}[t] e^{-\imagunit \nu \frac{2\pi}{S} t}.
\end{align}
After removing the cyclic prefix and after taking DFT, the received signal at the UEs and on the $\nu$th subcarrier can be written as
\begin{equation}
    \overline{\vect{y}}[\nu] = \overline{\vect{H}}[\nu] \overline{\vect{x}}[\nu] + \overline{\vect{w}}[\nu]
    \label{eq:rx-signal}
\end{equation}
for \( \nu = 0, \ldots, S-1 \). Here, \(\overline{\vect{x}}[\nu]\), \(\overline{\vect{y}}[\nu]\), and \(\overline{\vect{w}}[\nu]\) correspond to the \(\nu\)th column of 
\(\overline{\vect{X}}\), \(\overline{\vect{Y}}\), and \(\overline{\vect{W}}\), respectively. Moreover, we have white Gaussian noise $\overline{\vect{w}}[\nu]\sim \CN(\vect{0},\sigma^2\vect{I}_K)$.

We let $\mathcal{I} \subset \{0,\ldots,S-1\}$ denote the set of occupied subcarriers with $|\mathcal{I}| = S_{\mathcal{I}}$,
and let $\mathcal{G}$ denote the set of guard subcarriers.
For subcarrier $\nu \in \mathcal{I}$, the transmitted data symbol vector is
\begin{equation}
\vect{s}[{\nu}] = [s_{1}[\nu],\ldots,s_{K}[\nu]]^{\Ttran} \in \mathbb{C}^K,
\end{equation}
where $s_{k}[\nu]$ is drawn from a symbol constellation.
For $\nu \in \mathcal{G}$, we set $\vect{s}[{\nu}] = \vect{0}$.

 The nonlinear precoder is designed such that each UE forms a soft estimate by scaling its received signal as
$\beta\,\overline{\vect{y}}[\nu]$, where $\beta>0$ denotes a positive scaling factor.
Our objective is to design the phase-quantized CE precoding vectors $\vect{x}[t]$
so as to minimize the distortion, i.e., the average distance between the intended symbols and the scaled received signal, i.e.,
\begin{align}
   \sum_{\nu\in \mathcal{I}} \mathbb{E} \left\{\left\Vert \beta\overline{\vect{y}}[\nu]-\vect{s}[\nu]\right\Vert^2\right\} \label{eq:expected-error}.
\end{align}
 The expectation in \eqref{eq:expected-error} with respect to the random noise realizations can be expressed as
\begin{align}
   \sum_{\nu \in \mathcal{I}} \left \Vert  \vect{s}[{\nu}]-\beta\overline{\vect{H}}[\nu]\overline{\vect{x}}[\nu]\right\Vert^2+\sigma^2 S_{\mathcal{I}}K\beta^2.
\end{align}

\section{CE Quantized Precoding with Power Control}

The entries of the time-domain precoding vectors $\vect{x}[n]$ are constrained to belong to a
phase-quantized CE alphabet due to the use of low-resolution DACs and highly
power-efficient nonlinear power amplifiers.  We present two schemes with increasing flexibility:
(i) a classical maximum-power baseline,
and (ii) power control across APs. The second scheme is solved using an iterative alternating procedure in which the CE precoding
and the power control coefficients are updated in turn.

\subsection{Classical Approach with Maximum Antenna Power}

In the classical baseline, all antennas transmit with the maximum CE power and no
power control is performed across APs.
Thus, the transmit samples satisfy $x_{m}[n]\in\mathcal{X}_p$ for all antennas, where the feasible set is given by
\begin{equation}
\mathcal{X}_p \triangleq
\left\{
\sqrt{P_{\rm ant}} e^{\imagunit \frac{\pi(2q+1)}{2^p}}
\,:\,
q=0,\ldots,2^p-1
\right\},
\end{equation}
for $p$-bit DAC quantization, where $P_{\rm ant}$ is the maximum antenna power.

The CE precoding problem is
\begin{equation}\label{eq:prob_classical}
\minimize{\{\overline{\vect{x}}[\nu]\},\beta\geq 0}
\sum_{\nu \in \mathcal{I}}
\left\|
\vect{s}[\nu] - \beta \overline{\vect{H}}[\nu]\overline{\vect{x}}[\nu]
\right\|^2+\sigma^2 S_{\mathcal{I}}K\beta^2,
\end{equation}
together with $x_{m}[n]\in\mathcal{X}_p$. Following the SQUID-OFDM framework in \cite{jacobsson2018nonlinear},
we reformulate the classical CE precoding problem
by absorbing the receive-side scaling factor $\beta$ into the transmit
variables.
Specifically, we define the scaled frequency-domain precoding vectors as
\begin{equation}
\overline{\vect{b}}[\nu] \triangleq \beta\,\overline{\vect{x}}[\nu],
\end{equation}
and equivalently denote by
\(
\overline{\vect{B}} = [\overline{\vect{b}}[0],\ldots,\overline{\vect{b}}[S-1]]
\)
the scaled frequency-domain precoding matrix. Then, the sum-mean square error (MSE) objective in
\eqref{eq:prob_classical} can be written as
\begin{align}
\sum_{\nu\in\mathcal{I}}
\big\|
\vect{s}[\nu]-\overline{\vect{H}}[\nu]\overline{\vect{b}}[\nu]
\big\|^2
+ \sigma^2 S_{\mathcal{I}}K\beta^2 .
\end{align}

For CE signaling with maximum antenna power,
the time-domain precoding matrix satisfies
\begin{equation}
\big\|\mathrm{vec}(\vect{X})\big\|_\infty^2 = P_{\rm ant},
\end{equation}
which implies
\begin{equation}
\big\|\mathrm{vec}(\overline{\vect{B}} \vect{F}_S^{\Htran})\big\|_\infty^2
= \beta^2 P_{\rm ant}.
\end{equation}
Using this identity, the precoding problem
can be written by relaxing CE quantization constraints as 
\begin{align}\label{eq:pp_classical}
\minimize{\overline{\vect{B}}}
\quad &
\sum_{\nu\in\mathcal{I}}
\big\|
\vect{s}[\nu]-\overline{\vect{H}}[\nu]\overline{\vect{b}}[\nu]
\big\|^2
+ \gamma
\big\|
\mathrm{vec}(\overline{\vect{B}} \vect{F}_S^{\Htran})
\big\|_\infty^2,
\end{align}
where $\gamma \triangleq \sigma^2 S_{\mathcal{I}}K/P_{\rm ant}$.

Problem \eqref{eq:pp_classical} corresponds to the
$\ell_2$--$\ell_\infty$ relaxed formulation of the
CE precoding problem and is convex. \eqref{eq:pp_classical} can be efficiently solved using
Douglas--Rachford splitting, leading to the SQUID-OFDM iterations
in \cite{jacobsson2018nonlinear}.
It constitutes the OFDM extension of the SQUID precoder
originally proposed for frequency-flat channels \cite{Jacobsson2017a}.

Once \eqref{eq:pp_classical} is solved, the corresponding
time-domain waveform is obtained as
\begin{equation}
\vect{B}^\star = \overline{\vect{B}}^\star \vect{F}_S^{\Htran}.
\end{equation}
Finally, each entry of $\vect{B}^\star$ is mapped element-wise onto the discrete phase-quantized CE alphabet $\mathcal{X}_p$.

\subsection{Power Control Across APs}

We next introduce power control across APs, where the same power coefficients are applied
throughout $S$ time instants in an OFDM symbol.
Let $p_l\ge 0$ denote the power coefficient of AP $l$.
We define the diagonal scaling matrix
\begin{align}
\vect{P} \triangleq \diag(\boldsymbol{\rho} \kron \vect{1}_N),
\end{align}
where $\boldsymbol{\rho}=[\rho_1,\ldots,\rho_L]^{\Ttran}$ with $\rho_l=\sqrt{p_l}$.
Each AP is subject to an individual power constraint, which implies the box constraint
\begin{equation}
0 \le \rho_l \le \sqrt{P_{\rm ant}}, \qquad \forall l.
\end{equation}

The received signal model becomes
\begin{equation}
\overline{\vect{y}}[\nu]
=
\overline{\vect{H}}[\nu]\vect{P}\overline{\vect{x}}[\nu]
+
\overline{\vect{w}}[\nu].
\end{equation}
We aim to minimize the distortion under per-AP power limits:
\begin{subequations} \label{eq:prob_realPA}
\begin{align}
\minimize{\{\rho_l\},\{\overline{\vect{x}}[\nu]\},\beta\geq 0}
\quad &
 \sum_{\nu \in \mathcal{I}}
\left\|
\vect{s}[\nu]
-
\beta \overline{\vect{H}}[\nu]\vect{P}\overline{\vect{x}}[\nu]
\right\|^2 
+\sigma^2 S_{\mathcal{I}}K\beta^2
\\
\text{subject to}\quad &
0 \le \rho_l \le \sqrt{P_{\rm ant}},\quad \forall l,
\end{align}
\end{subequations}
together with $x_{m}[n]$ belonging to discrete phase-quantized CE alphabet.
We solve \eqref{eq:prob_realPA} via alternating optimization:
for fixed $\{\rho_l\}$ we update $\{\overline{\vect{x}}[\nu]\}$ via SQUID-OFDM. For fixed precoding vectors $\{\overline{\vect{x}}[\nu]\}$, the objective is a convex quadratic function of $\beta\geq 0$.
Define the noiseless received vector
\begin{equation}
\vect{u}[\nu] \triangleq \overline{\vect{H}}[\nu]\vect{P}\overline{\vect{x}}[\nu]\in\mathbb{C}^{K}.
\end{equation}
Then, by expanding the MSE term and using $\mathbb{E}\{\|\beta \overline{\vect{w}}[\nu]\|^2\}
= \beta^2\sigma^2 K$, the $\beta$-dependent part of the distortion can be written as
\begin{align}
J(\beta)
&=
\sum_{\nu\in\mathcal{I}}
\left\|
\vect{s}[\nu]-\beta \vect{u}[\nu]
\right\|^2
+\sigma^2 S_{\mathcal{I}}K\,\beta^2 \nonumber \\
&=
a\,\beta^2-2b\,\beta+\text{const},
\end{align}
where
\begin{align}
a &\triangleq \sum_{\nu\in\mathcal{I}}\|\vect{u}[\nu]\|^2
+\sigma^2 S_{\mathcal{I}}K, \\
b &\triangleq \sum_{\nu\in\mathcal{I}}
\Re\!\left(\vect{u}^{\Htran}[\nu]\vect{s}[\nu]\right).
\end{align}
Since $a>0$, the unconstrained minimizer is $\beta^{\rm unc}=b/a$. Enforcing $\beta\ge 0$ yields the
closed-form optimal update
\begin{equation}
\beta^\star
=
\max\left\{0,\ \frac{b}{a}
\right\}.
\label{eq:beta_closed_form}
\end{equation}
In our alternating procedure, we update $\beta$ using \eqref{eq:beta_closed_form} after each precoder update.
Next, for fixed $\{\overline{\vect{x}}[\nu]\}$ and $\beta$ we update each $\rho_l$ in sequel in closed form.

\subsubsection*{Closed-form per-AP update of power control coefficients}
Fix $\{\overline{\vect{x}}[\nu]\}$ and $\beta>0$.
For each subcarrier $\nu$, we partition the frequency-domain channel as
\begin{align}
\overline{\vect{H}}[\nu]=\big[\overline{\vect{H}}_{1}[\nu],\ldots,\overline{\vect{H}}_{L}[\nu]\big],
\qquad \overline{\vect{H}}_{l}[\nu]\in \mathbb{C}^{K\times N},
\end{align}
and similarly partition the precoding vector as
\begin{align}
\overline{\vect{x}}[\nu]=\big[\overline{\vect{x}}_{1}^{\Ttran}[\nu],\ldots,\overline{\vect{x}}_{L}^{\Ttran}[\nu]\big]^{\Ttran},
\qquad \overline{\vect{x}}_{l}[\nu]\in\mathbb{C}^{N}.
\end{align}
Then the noiseless received vector is
\begin{align}
\beta\,\overline{\vect{H}}[\nu]\vect{P}\overline{\vect{x}}[\nu]
=\beta \sum_{l=1}^L \rho_l \,\overline{\vect{H}}_l[\nu]\,\overline{\vect{x}}_{l}[\nu].
\end{align}
Define the AP-$l$ contribution vector
\begin{align}
\vect{a}_{\nu,l}\triangleq \overline{\vect{H}}_l[\nu]\overline{\vect{x}}_{l}[\nu]\in \mathbb{C}^{K}.
\end{align}
Moreover, when updating $\rho_l$, define the residual that excludes AP $l$ as
\begin{align}
\vect{e}_{\nu}^{(-l)} \triangleq \vect{s}[\nu]
-\beta\sum_{j\neq l}\rho_j\,\vect{a}_{\nu,j}.
\end{align}
The distortion term for given $\nu$ becomes
\[
\left\|\vect{s}[\nu]-\beta\sum_{j=1}^L \rho_j\vect{a}_{\nu,j}\right\|^2
=\big\|\vect{e}_{\nu}^{(-l)}-\beta\rho_l \vect{a}_{\nu,l}\big\|^2,
\]
and hence the $\rho_l$-dependent part of the total objective is a convex quadratic:
\begin{align}
\sum_{\nu\in\mathcal{I}}
\big\|\vect{e}_{\nu}^{(-l)}-\beta\rho_l \vect{a}_{\nu,l}\big\|^2
&=
A_l \rho_l^2 -2B_l\rho_l + \text{const},
\end{align}
where
\begin{align}
A_l &\triangleq \beta^2 
\sum_{\nu\in\mathcal{I}} \|\vect{a}_{\nu,l}\|^2,\\
B_l &\triangleq \beta \sum_{\nu\in\mathcal{I}} 
\Re\left(\vect{a}_{\nu,l}^{\Htran}\vect{e}_{\nu}^{(-l)}\right).
\end{align}

Since the problem is subject only to the box constraint
$0\le \rho_l \le \sqrt{P_{\rm ant}}$,
each coefficient admits a closed-form update obtained by projecting the unconstrained minimizer onto the feasible interval.
The unconstrained optimum is
\begin{equation}
\rho_l^{\mathrm{unc}}=\frac{B_l}{A_l}.
\end{equation}
Hence, the optimal long-term coefficient is
\begin{align}
&\rho_l^\star
=
\min\!\left\{
\sqrt{P_{\rm ant}},
\;
\max\!\left\{0,\ \frac{B_l}{A_l}\right\}
\right\} \nonumber \\
&\Longrightarrow\qquad
p_l^\star = (\rho_l^\star)^2 .
\end{align}

In particular,
if $B_l \le 0$ then $\rho_l^\star=0$,
while if $B_l/A_l \ge \sqrt{P_{\rm ant}}$ the per-AP power constraint becomes active.

\paragraph*{Integration with SQUID-OFDM under per-AP power control}

When the power control coefficients $\{\rho_l\}$ are given,
the CE precoder is updated using the same SQUID-OFDM solver
as in the classical case, but with an effective channel that absorbs the
power control matrix.
Specifically, defining the diagonal scaling matrix
$\vect{P}=\diag(\boldsymbol{\rho}\kron\vect{1}_N)$, the received signal model becomes
\begin{equation}
\overline{\vect{y}}[\nu]
=
\overline{\vect{H}}[\nu]\vect{P}\overline{\vect{x}}[\nu]
+
\overline{\vect{w}}[\nu].
\end{equation}
Instead of modifying the SQUID algorithm itself, we incorporate the power
coefficients into the channel and define the effective frequency-domain channel
\begin{equation}
\widetilde{\vect{H}}[\nu] \triangleq \overline{\vect{H}}[\nu]\vect{P}.
\end{equation}
Then, for fixed $\{\rho_l\}$, the precoding update reduces to solving
\eqref{eq:pp_classical} with $\overline{\vect{H}}[\nu]$ replaced by
$\widetilde{\vect{H}}[\nu]$.

Importantly, the SQUID-OFDM solver is always applied under the normalized
CE constraint with $P_{\rm ant}=1$. The procedure described above is repeated for several iterations, where all variables are updated in an alternating manner.

\section{Numerical Results}

In this section, we evaluate the performance of the proposed CE quantized precoding with AP-level power control for the downlink of a cell-free massive MIMO--OFDM system. Since the proposed precoder operates under phase-quantized CE constraints in the symbol level, we assess the communication performance in terms of the uncoded BER. We compare the proposed power-control schemes with a classical CE baseline that transmits with maximum antenna power and performs no power adaptation across APs.

We consider a cell-free deployment with $L=49$ APs, each equipped with $N=4$ antennas, serving $K=25$ single-antenna UEs over a square area with side length $350$\,m. The APs are placed on a uniform grid, while the UEs are dropped uniformly at random. The height difference between an AP and a UE is $10$\,m. OFDM transmission with $S=2000$ subcarriers and subcarrier spacing $\Delta f = 15$\,kHz is employed, yielding a system bandwidth of $B = S\Delta f$. Among the $S$ subcarriers, $S_{\mathcal{I}}=1200$ are occupied for data transmission, and QPSK modulation is used. The indices of the occupied subcarriers are $\{1,\ldots,600\}$ and $\{1400,\ldots,1999\}$.

The large-scale fading is generated according to the pathloss--shadowing model
\begin{equation}
\beta_{kl}[{\mathrm{dB}}] = -30.5 - 36.7\log_{10}(d_{kl}) + z_{kl},
\end{equation}
where $d_{kl}$ denotes the 3D distance (in meters) between AP $l$ and UE $k$, and $z_{kl}\sim\mathcal{N}(0,4^2)$ models log-normal shadow fading. The noise power is computed as
\begin{equation}
\sigma^2[{\mathrm{dB}}] = -204 + 10\log_{10}(B) + \mathrm{NF},
\end{equation}
with noise figure $\mathrm{NF}=5$\,dB. Spatial correlation matrices are generated using the local scattering model with Gaussian azimuth and elevation angular spreads of $15^\circ$, and half-wavelength antenna spacing \cite{cell-free-book}. The results are obtained over $300$ independent setups.

The small-scale fading is modeled by a frequency-selective tapped-delay-line channel with $T+1=5$ taps. For each setup, we generate a random power-delay profile $\{\tilde{p}_t\}_{t=0}^{T}$ by drawing $T$ independent and identically distributed uniform random variables on $(0,1)$, sorting them in descending order, and normalizing them such that $\sum_{t=0}^{T} \tilde{p}_t = 1$. Conditioned on $\{\tilde{p}_t\}$, the $t$th tap between AP $l$ and UE $k$ is generated as a spatially correlated Rayleigh fading vector with covariance $\tilde{p}_t \mathbf{R}_{kl}$, i.e.,
\begin{equation}
\mathbf{h}_{kl}[t] \sim \mathcal{CN}\!\left(\mathbf{0},\, \tilde{p}_t \mathbf{R}_{kl}\right), \quad t=0,\ldots,T,
\end{equation}
where $\mathbf{R}_{kl}\in\mathbb{C}^{N\times N}$ is obtained from the local scattering model described above.

Fig.~1 illustrates a representative network geometry for a randomly generated setup, together with the AP-level transmit power distribution obtained from the proposed power-control algorithm with a 2-bit DAC. The APs are positioned on a uniform grid, while the UEs are randomly dropped within the coverage area. The color associated with each AP indicates its transmit power relative to the maximum allowable power, where warmer colors correspond to higher transmit power levels.

The figure illustrates that the resulting power distribution is highly non-trivial and does not follow a simple distance-based pattern. In particular, some APs located close to multiple UEs are assigned reduced transmit power levels, indicating that the proposed symbol-level CE power-control mechanism accounts for multiuser interference effects rather than merely amplifying the strongest links. Consequently, the transmit power distribution is spatially heterogeneous, reflecting a joint adaptation to channel conditions and UE geometry under CE quantization constraints.

\begin{figure}[t!]
		\vspace{0.1cm}
	\begin{center}
		\includegraphics[trim={0.6cm 0cm 1cm 0.6cm},clip,width=8.1cm]{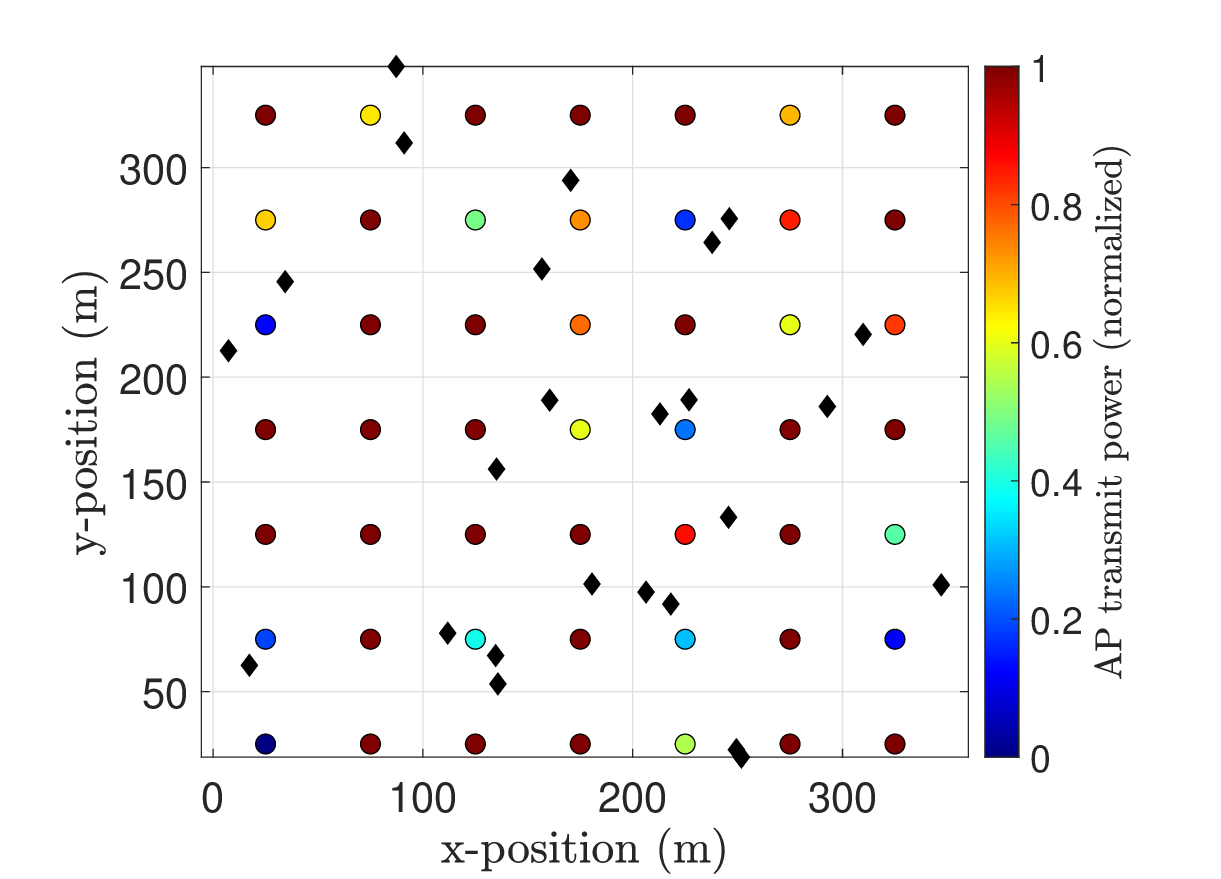}
		\caption{Example network geometry. Colored circles denote APs, where the color represents the normalized transmit power, while black diamonds indicate UE locations. The figure illustrates the spatial power distribution achieved by the proposed CE power-control scheme.}
\label{fig:sim3}
\vspace{-3mm}
	\end{center}
\end{figure}

Fig.~2 presents the uncoded BER performance for different DAC resolutions and power-control strategies. The results include both $1$-bit and $2$-bit phase-quantized CE precoding, where each case is evaluated with a maximum-power baseline and the proposed AP-level power-control scheme. The horizontal axis corresponds to the UE indices sorted in descending order of BER for each setup; hence, the left side of the figure represents the most unfavorable UEs, while the right side corresponds to more favorable channel conditions.

The results reveal that the proposed power-control mechanism substantially reduces the uncoded BER across the entire UE population. In particular, significant improvements are observed for the worst-performing UEs, indicating that power adaptation effectively mitigates unfavorable interference and channel conditions. Increasing the DAC resolution from $1$ bit to $2$ bits further enhances the performance gains achieved by power control, highlighting that power control is a critical component of CE quantized precoding in cell-free massive MIMO systems.

\begin{figure}[t!]
		\vspace{0.1cm}
	\begin{center}
		\includegraphics[trim={0cm 0cm 1cm 0.6cm},clip,width=8.1cm]{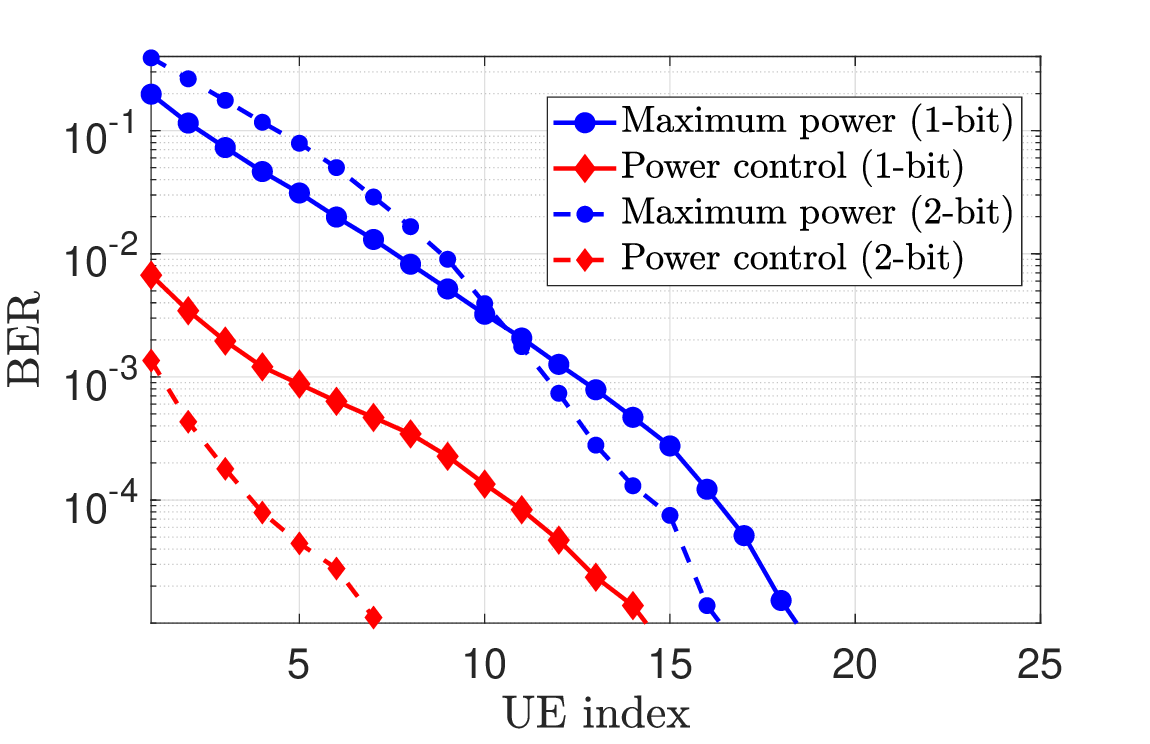}
		\caption{Uncoded BER performance versus UE index for $1$-bit and $2$-bit phase-quantized CE precoding. The curves include the maximum-power baseline and the proposed AP-level power-control scheme. 
}
\vspace{-3mm}
\label{fig:sim3}
	\end{center}
\end{figure}

\section{Conclusion}

In this paper, we investigated symbol-level CE quantized precoding with AP-level power control for the downlink of cell-free massive MIMO--OFDM systems. The proposed framework integrates adaptive power control into time-domain CE waveform design, enabling flexible transmit power adaptation across distributed APs while preserving strict phase-quantized constraints. Numerical results demonstrated that the proposed power-control strategy significantly improves the uncoded BER performance compared to conventional maximum-power transmission, with additional gains observed when increasing the DAC resolution. Moreover, since the power-control coefficients evolve on a slower channel timescale, the required control signaling between the central processing unit and the APs remains limited, making the approach attractive for practical deployments with constrained fronthaul resources.

\bibliographystyle{IEEEtran}
\bibliography{IEEEabrv,refs}

\end{document}